%% file: main.tex
\documentclass[reprint,
superscriptaddress,amsmath,amssymb,aps, longbibliography
]{revtex4-2}

\usepackage{graphicx}
\usepackage{dcolumn}
\usepackage{bm}
\usepackage{siunitx}
\usepackage[noend]{algorithmic}
\usepackage{algorithm}
\usepackage[skins]{tcolorbox}
\usepackage{setspace}
\usepackage{blkarray}

\usepackage{xcolor}
\usepackage{soul}
\usepackage{systeme}
\usepackage{footmisc}

\usepackage{scalerel} 

\makeatletter
\def\maketag@@@#1{\hbox{\m@th\normalfont\normalsize#1}}
\makeatother

\newcommand{\blue}[1]{\textcolor{black}{#1}}

\usepackage{booktabs}
\usepackage{hyperref}
\usepackage[capitalize]{cleveref}
\usepackage[utf8]{inputenc}

\usepackage[lining,semibold]{libertine} 
\usepackage{amsthm}
\usepackage[libertine, cmintegrals, bigdelims, vvarbb]{newtxmath}

\usepackage{amsmath}
\usepackage{amsfonts}
\usepackage{mathrsfs}
\usepackage{gensymb}
\usepackage{bbm}
\usepackage{dsfont}

\usepackage{pgfplots}
\usepgfplotslibrary{colormaps}

\usepackage{kbordermatrix}
\renewcommand{\kbldelim}{(}
\renewcommand{\kbrdelim}{)}

\usepackage{chemformula}
\usepackage[caption=false]{subfig}

\usepackage{soul}
\usepackage{xcolor}

\theoremstyle{definition}

\usepackage{scalerel} 

\definecolor{webgreen}{rgb}{0,.5,0}
\definecolor{webbrown}{rgb}{.6,0,0}
\definecolor{grigio}{rgb}{.85,.85,.85} 
\definecolor{RoyalBlue}{rgb}{0.0, 0.14, 0.4}
\definecolor{skyblue1}{rgb}{0.45,0.62,0.81}
\definecolor{skyblue2}{rgb}{0.2,0.39,0.64}
\definecolor{skyblue3}{rgb}{0.13,0.29,0.53}
\definecolor{scarlet1}{rgb}{0.93,0.16,0.16}
\definecolor{scarlet2}{rgb}{0.8,0,0}
\definecolor{scarlet3}{rgb}{0.64,0,0}

\definecolor{g}{gray}{0.50}

\hypersetup{%
    colorlinks=true, linktocpage=true, pdfstartpage=1, pdfstartview=FitV,%
    breaklinks=true, pdfpagemode=UseNone, pageanchor=true, pdfpagemode=UseOutlines,%
    plainpages=false, bookmarksnumbered, bookmarksopen=true, bookmarksopenlevel=1,%
    hypertexnames=true, pdfhighlight=/O,
    urlcolor=webbrown, linkcolor=RoyalBlue, citecolor=webgreen, 
    pdftitle={},%
    pdfauthor={Timur Aslyamov},%
    pdfsubject={},%
    pdfkeywords={},%
    pdfcreator={pdfLaTeX},%
    pdfproducer={LaTeX REVTeX}%
}

\begin{document}
\title{Nonequilibrium Fluctuation-Response Relations: From Identities to Bounds}

\author{Timur Aslyamov}
\email{timur.aslyamov@uni.lu}
\affiliation{Complex Systems and Statistical Mechanics, Department of Physics and Materials Science, University of Luxembourg, 30 Avenue des Hauts-Fourneaux, L-4362 Esch-sur-Alzette, Luxembourg}

\author{Krzysztof Ptaszy\'{n}ski}
\email{krzysztof.ptaszynski@ifmpan.poznan.pl}
\affiliation{Institute of Molecular Physics, Polish Academy of Sciences, Mariana Smoluchowskiego 17, 60-179 Pozna\'{n}, Poland}

\author{Massimiliano Esposito}
\email{massimiliano.esposito@uni.lu}
\affiliation{Complex Systems and Statistical Mechanics, Department of Physics and Materials Science, University of Luxembourg, 30 Avenue des Hauts-Fourneaux, L-4362 Esch-sur-Alzette, Luxembourg}

\date{\today}

\begin{abstract}
In nonequilibrium steady states of Markov jump processes, we derive exact Fluctuation-Response Relations (FRRs) that express the covariance between any pair of currents in terms of static responses in a notably simple form, thus generalizing the fluctuation-dissipation theorem far from equilibrium.
We begin by considering perturbations in the symmetric part of the rates. We demonstrate that FRRs imply a hierarchy of thermodynamic bounds. These hierarchies prove the recently conjectured Response Thermodynamic Uncertainty Relation (R-TUR), which bounds the ratio between any current's response and its variance by the entropy production rate (EPR). We furthermore strengthen this bound in two distinct ways, using partial EPR in one case and pseudo-EPR in the other.
For perturbations in the antisymmetric part of the rates, we show that the ratio between any current's response and its variance is bounded by traffic, a metric representing the total number of transitions per unit time in the system.
As an application, we use FRRs to explain the origin of positive correlations between currents in Coulomb-blockaded systems previously observed in experiments.
\end{abstract}

\maketitle
\textit{Introduction---}Fluctuations in Nonequilibrium Steady States (NESS) of Markov jump processes play a central role in nonequilibrium physics.
Close-to-equilibrium, the seminal fluctuation-dissipation theorem, links equilibrium fluctuations to the linear response of a system to external forces~\cite{kubo1966fluctuation,kubo2012statistical,stratonovich2012nonlinear,marconi2008fluctuation,forastiere2022linear}. 
This result has been generalized by relating fluctuations to the linear response in a NESS~\cite{agarwal1972fluctuation,seifert2010fluctuation,prost2009generalized,altaner2016fluctuation,chun2021nonequilibrium,baiesi2009fluctuations,shiraishi2023introduction}. 

Another, apparently unrelated, result regarding nonequilibrium fluctuations are Thermodynamic Uncertainty Relations (TURs). The original TUR upper bounds the squared precision of current observables (i.e., the ratio between the square of the average current $\mathcal{J}$ and its variance $\langle\langle\mathcal{J}\rangle\rangle$) by half of the Entropy Production Rate (EPR) $\dot{\sigma}$, where the Boltzmann constant $k_B=1$ 
\cite{barato2015thermodynamic,gingrich2016dissipation,pietzonka2016universal, pietzonka2017finite,horowitz2017proof,falasco2020unifying, horowitz2020thermodynamic,vu2020entropy,dechant2020fluctuation,dechant2019arxiv,van2023thermodynamic}. A tighter version also replaces $\dot{\sigma}/2$ by the pseudo-EPR \cite{shiraishi2021optimal}.  
\blue{Another inequality, Kinetic Uncertainty Relation (KUR),} bounds the squared precision by traffic, a quantity that measures the total number of transitions per unit time in the system \cite{di2018kinetic}.

A first connection between response and TUR was obtained in \cite{dechant2020fluctuation,dechant2019arxiv} by upper bounding the ratio of the squared difference between a perturbed and unperturbed average observable to its unperturbed variance by twice the Kullback-Leibler divergence between the perturbed and unperturbed path probability taken by the system. This bound was tightened in \cite{falasco2022beyond} and asymmetric response relations were also obtained.
Recently, a more explicit connection was found with Response-TUR (R-TUR) conjectured in \cite{ptaszynski2024dissipation}, but it has not been proven so far.
This result upper bounds the precision of kinetic perturbations to currents (i.e., the ratio between the squared static response of a current to perturbations of kinetic barriers and the current variance) by half the EPR multiplied by the squared maximal rate of change of kinetic barriers in the system.
Thus, it relates response, fluctuations, and dissipation in the form of a bound, raising the important question of whether an exact relation exists between these quantities.

In this Letter, we derive such exact relations and coin them Fluctuation-Response Relations (FRRs).
They relate the covariance between two arbitrary currents to their static responses to arbitrary perturbations of the transition rate matrix. 
Hierarchies of thermodynamic bounds can be derived from them. 
For symmetric perturbations of the rates, we not only prove the R-TUR conjectured in \cite{ptaszynski2024dissipation}, but also tighten them in terms of partial-EPR and pseudo-EPR. 
For antisymmetric perturbations of the rates, we find a related hierarchy of \blue{Response-KUR (R-KUR) bounds}, but this time in terms of traffic rather than entropy production. 
We also show that FRRs straightforwardly imply the original TUR as well its pseudo-EPR and kinetic versions. 
Finally, we apply FRRs to explain the origin of positive correlations in the currents across a double quantum dot setup measured in the Coulomb-blockade regime.

\textit{Setup---}We consider a continuous-time Markov jump process among $N$ discrete states. In NESS we have
\begin{align}
\label{eq:NESS}
    d_t\boldsymbol{\pi} = \mathbb{W}\cdot\boldsymbol{\pi} = 0 \,,
\end{align}
where $\boldsymbol{\pi}=(\dots,\pi_n,\dots)^\intercal$ is the vector of state probabilities $\pi_n$ with $\sum_n\pi_n=1$. The matrix $\mathbb{W}$ is the rate matrix with off-diagonal elements $W_{nm}=\sum_{e}[ W_{+e}\delta_{s(+e)m}\delta_{t(+e)n} + W_{-e}\delta_{s(+e)n}\delta_{t(+e)m}]$, 
where $s(\pm e)$ is the source of edge $\pm e$, and $t(\pm e)$ is the target of edge $\pm e$; and diagonal elements $W_{nn}=-\sum_{m \neq n}W_{mn}$. Every transition rate can be generically parameterized as
\begin{align}
\label{eq:rates-model}
    W_{\pm e}=\exp\big[B_e\pm S_e/2\big]\,,
\end{align}
where $B_e$ and $S_e$ parameterize the symmetric and antisymmetric part of the transition rate, respectively. 
For physical systems in contact with thermal reservoirs (rates that satisfy the local detailed balance), the term $B_e$, characterizes the kinetic barriers \blue{which can be controlled by varying catalyst concentrations (e.g., enzymes)~\cite{Wachtel_2018}, applying magnetic fields (e.g., via the radical pair mechanism in magnetoreception)~\cite{ritz2000model, hore2016radical, wiltschko2019magnetoreception, zadeh2022magnetic}, adjustment of tunnel barriers~\cite{gustavsson2006counting,gustavsson2009electron,sanchez2019Autonomous} or potential barriers~\cite{freitas2021stochastic, gopal2022Large} by gate voltages in nano-electronics}; while the term $S_e$, is the change in entropy in the reservoir due to a transition $e$ that includes changes in thermodynamic forces and energy landscape \cite{rao2018conservation, falasco2023macroscopic, owen2020universal}.
The current through the edge $e$ is $j_e = W_{+e}\pi_{s(+e)}-W_{-e}\pi_{t(+e)}$; all unoriented transitions through the edge $e$ are characterized by a traffic $\tau_e= W_{+e}\pi_{s(+e)}+W_{-e}\pi_{t(+e)}$. 
The dissipation in the system is defined by the total EPR, $\dot{\sigma}=\sum_e\dot{\sigma}_e$, where $\dot{\sigma}_e =j_e\ln(W_{+e}\pi_{s(+e)}/W_{-e}\pi_{t(+e)})$ is the single edge EPR.
A steady state is an equilibrium if $\dot{\sigma}=0$ (and thus all $j_e=0$) and a NESS if $\dot{\sigma} > 0$.

To characterize the average and fluctuations of current observables, we introduce the stochastic counting variable $k_e(t)=k_{+e}(t)-k_{-e}(t)$, where $k_{\pm e}(t)$ is the number of jumps in the direction $\pm e$ during the time interval $[0,t]$. The average of an arbitrary current observable $\mathcal{J}$ is given by
\begin{align}
\mathcal{J} \equiv \lim_{t \rightarrow \infty} t^{-1} \left \langle \sum_e x_e k_e(t) \right \rangle= \sum_e x_e j_e\,,
\end{align}
where $\boldsymbol{x}=(\dots, x_e, \dots)^\intercal$ is an arbitrary vector. The covariance between two arbitrary current observables $\langle\langle\mathcal{J},\mathcal{J}'\rangle\rangle$, where $\mathcal{J}'=\sum_{e}x'_e j_e$ with $\boldsymbol{x}'=(\dots, x_e', \dots)^\intercal$, is in turn
\begin{align}
\label{eq:cov-def}
\langle\langle\mathcal{J},\mathcal{J}'\rangle\rangle \equiv \lim_{t \rightarrow \infty} t^{-1} \left \langle \sum_{e} x_e \Delta k_e(t) \sum_{e'} x_{e'} \Delta k_{e'}(t) \right \rangle=\boldsymbol{x}^\intercal \mathbb{C} \boldsymbol{x}'\,,
\end{align}
where $\Delta k_e(t)=k_e(t)-\langle k_e(t) \rangle$ and $\mathbb{C}$ is the covariance matrix of edge currents with elements $C_{ee'}=\lim_{t \rightarrow \infty} t^{-1} \langle \Delta k_e(t) \Delta k_{e'}(t) \rangle$.
The exact expression for the covariance matrix $\mathbb{C}$ is given in \cref{eq:matC} in \cref{sec:derivation_FRR}.

\textit{Fluctuation-Response Relations---}The FRRs are derived in detail in \cref{sec:derivation_FRR}. They link the covariance of currents to the static response of currents to perturbations in the rates. 
\blue{
The term ``static response'' means that one considers the linear response of the steady-state currents to the perturbation of the rates \cite{aslyamov2024general}. From an operational perspective, it corresponds to measuring the current responses for a sufficiently long time after introducing the perturbation of the rates, so that the system has relaxed to the new stationary state.
In the main text, we assume that the vector $\boldsymbol{x}$ does not depend on the perturbation parameter, resulting in static responses of the current observable $\mathcal{J}$ as $d_{B_{e}}\mathcal{J} = \sum_{e'} x_{e'} d_{B_{e}} j_{e'}$ and $d_{S_{e}}\mathcal{J} = \sum_{e'} x_{e'} d_{S_{e}} j_{e'}$, where $d_{p}j_{e}(p,\boldsymbol{\pi}(p)) = \partial_{p} j_{e} + \sum_{n}\partial_{\pi_n}j_{e}\partial_{p}\pi_n$, with $p\in\{B_{e},S_{e}\}$. 
The results for an arbitrary $\boldsymbol{x}$ are shown in \cref{sec:derivation_FRR}. 
}

For symmetric perturbations, $B_e$, the FRR reads
\begin{align}
\label{eq:covar-exact-sym}
 \langle\langle\mathcal{J},\mathcal{J}'\rangle\rangle = \nabla_{B}^\intercal\mathcal{J}\cdot\mathbb{D}^{-1}\cdot\nabla_{B}\mathcal{J}' = \sum_{e}\frac{\tau_e}{j_e^2}d_{B_e}\mathcal{J}d_{B_e}\mathcal{J}'\,, 
\end{align}
here, $\nabla_{B}\mathcal{J}=(\dots, d_{B_e}\mathcal{J}, \dots)^\intercal$
and $\mathbb{D}=\text{diag}(\dots,j_e^2/\tau_e,\dots)$ is the diagonal matrix with the pseudo-EPR as trace, $\dot{\Pi}=\text{tr}\,\mathbb{D}=\sum_{e}j_e^2/\tau_e$. 
The pseudo-EPR is a measure of irreversibility (nonnegative and zero at equilibrium). It coincides with $\dot{\sigma}/2$ close to equilibrium [$\dot{\sigma}_e/2=j_e\text{arctanh}(j_e/\tau_e)=j_e^2/\tau_e+\mathcal{O}(j_e^4)$] and plays an important role in the generalization of TURs \cite{dechant2021improving,shiraishi2021optimal,van2022unified,dechant2022minimum}.
The current variance $\langle\langle\mathcal{J}\rangle\rangle$ is obtained from \cref{eq:covar-exact-sym} when $\mathcal{J}=\mathcal{J}'$ as
\begin{align}
\label{eq:var-exact-sym}
 \langle\langle\mathcal{J}\rangle\rangle  = \sum_{e} \frac{\tau_e}{j_e^2}\big(d_{B_e}\mathcal{J}\big)^2\,. 
\end{align}
For antisymmetric perturbations, $S_e$, the FRR for the covariance reads
\begin{align}
\label{eq:covar-exact-antisym}
    \langle\langle\mathcal{J},\mathcal{J}'\rangle\rangle &= 4\nabla_{S}^\intercal\mathcal{J}\cdot\mathbb{T}^{-1}\cdot\nabla_{S}\mathcal{J}' = \sum_{e} \frac{4}{\tau_e} d_{S_e}\mathcal{J}d_{S_e}\mathcal{J}'\,,
\end{align}
and the FRR for the variance is
\begin{align}
\label{eq:var-exact-antisym}
    \langle\langle\mathcal{J}\rangle\rangle &=\sum_{e} \frac{4}{\tau_e}\big(d_{S_e}\mathcal{J}\big)^2\,. 
\end{align}
Here, $\nabla_{S}\mathcal{J}=(\dots, d_{S_e}\mathcal{J}, \dots)^\intercal$, 
and $\mathbb{T}= \text{diag}(\dots,\tau_e,\dots)$ is a diagonal matrix with total traffic as trace, $\mathcal{T}=\text{tr}\,\mathbb{T} =\sum_e\tau_e$. 
Note that even for single edge currents, $\mathcal{J}=j_e$, the right-hand sides of \cref{eq:covar-exact-sym,eq:var-exact-sym,eq:covar-exact-antisym,eq:var-exact-antisym} include contributions from all edges of the network.
\blue{
We note that \cref{eq:covar-exact-sym,eq:var-exact-sym} still hold at a stalling edge where both the denominator, $j_e$, and the numerator, $\tau_e d_{B_e}\mathcal{J}d_{B_e}\mathcal{J}'$, tend to zero, because their ratio remains finite~\cite{aslyamov2024general}.  
}

\textit{FRRs close to equilibrium---}We now parameterize the antisymmetric part of \cref{eq:rates-model} as $S_e=F_e + E_{s(+e)}-E_{t(+e)}$, where $E_{n}$ can be interpreted as state energies and $F_e$ as nonconservative thermodynamic forces \cite{rao2018conservation}. At steady state $\dot{\sigma}= \boldsymbol{j}^\intercal\boldsymbol{F}$, where $\boldsymbol{F}=(\dots,F_e,\dots)^\intercal$. For $\boldsymbol{F} = \boldsymbol{0}$, the system steady state is an equilibrium state $\pi_n^\text{eq}=\exp(-E_n)/\sum_m\exp(-E_m)$. 
In the close-to-equilibrium regime, the forces $F_e$ are small. Currents can be linearly expanded in forces as $\boldsymbol{j}=\mathbb{L}\boldsymbol{F}$, where $\mathbb{L}=\mathbb{L}^\intercal$ is the positive semidefinite Onsager matrix; see the explicit form in \cite{vroylandt2019ordered} and \footnote{see Appendix B in \cite{ptaszynski2024dissipation}}. \blue{
Comparing, on one hand, the response calculated as $\mathcal{J} = \boldsymbol{x}^\intercal \mathbb{L} \boldsymbol{F} \rightarrow \nabla_S^\intercal \mathcal{J}|_{\boldsymbol{F}=0} = \boldsymbol{x}^\intercal \mathbb{L}$ and, on the other hand, the response calculated by \cref{eq:response-general} as $\nabla_S^\intercal \mathcal{J}|_{\boldsymbol{F}=0} = \tfrac{1}{2}\boldsymbol{x}^\intercal \mathbb{P}^\text{eq} \mathbb{T}_\text{eq}$, we find $\mathbb{L} = \tfrac{1}{2} \mathbb{P}_\text{eq} \mathbb{T}_\text{eq}$, where $\mathbb{P}_\text{eq}$ is the equilibrium projection matrix ($\mathbb{P}_\text{eq}^2 = \mathbb{P}_\text{eq}$); see \cref{eq:matP}. Inserting this expression for the response into \cref{eq:covar-exact-antisym}, we find
}
\begin{align}
\label{eq:FDT}    \langle\langle\mathcal{J}\rangle\rangle_\text{eq} = \boldsymbol{x}^\intercal\mathbb{P}_\text{eq}\mathbb{T}_\text{eq}\mathbb{P}_\text{eq}^\intercal\boldsymbol{x}=\boldsymbol{x}^\intercal\mathbb{P}_\text{eq}^2\mathbb{T}_\text{eq}\boldsymbol{x}=\boldsymbol{x}^\intercal2\mathbb{L}\boldsymbol{x}\,,
\end{align}
where we used $2\mathbb{L}=\mathbb{P}_\text{eq}\mathbb{T}_\text{eq}=\mathbb{T}_\text{eq}\mathbb{P}_\text{eq}^\intercal=2\mathbb{L}^\intercal$. 
From \cref{eq:FDT}, we find that equilibrium covariance matrix $\mathbb{C}_\text{eq}=2\mathbb{L}$, which is the fluctuation-dissipation theorem \cite{vroylandt2019ordered}. \blue{
Using $\boldsymbol{x}=\boldsymbol{F}$, we consider the EPR as the current observable $\dot{\sigma}=\boldsymbol{F}^\intercal \boldsymbol{j}=\boldsymbol{F}^\intercal \mathbb{L} \boldsymbol{F}$.
Inserting $\boldsymbol{x}=\boldsymbol{F}$ and $\mathcal{J} = \dot{\sigma}$ in \cref{eq:FDT}, we also find
$\langle\langle\dot{\sigma}\rangle\rangle_\text{eq}=2\boldsymbol{F}^\intercal\mathbb{L}\boldsymbol{F}=2\dot{\sigma}$ \cite{pietzonka2016universal}.
}
In Section SIII of \cite{SM} (\cref{sec:FRR-stalling}), we show that the FRR can also be used to derive the fluctuation-dissipation relation valid far-from-equilibrium but at stalling, originally derived in \cite{altaner2016fluctuation}.

\textit{Proving and generalizing R-TUR---}We now show how FRRs can be used to derive thermodynamic bounds.
We introduce the parameters $\varepsilon$ and $\eta$, which respectively control $B_e(\varepsilon)$ and $S_e(\eta)$, and the current responses
\begin{align}
    \label{eq:responses-eps-eta}
    d_\varepsilon\mathcal{J} = \sum_{e} b_e d_{B_e}\mathcal{J}\,,\quad  d_\eta\mathcal{J} = \sum_{e} z_e d_{S_e}\mathcal{J}\,,
\end{align}
with perturbation rates $b_e = \partial_\varepsilon B_e$ and $z_e = \partial_\eta S_e$. 

Two hierarchies of bounds are derived by combining \cref{eq:responses-eps-eta} with the FRRs \eqref{eq:var-exact-sym} and \eqref{eq:var-exact-antisym}, respectively. 
For the first hierarchy, we use Jensen's inequality to prove $2(a-b)^2/(a+b)\leq (a-b)\ln(a/b)$ which implies that $\tau_{e}/j_e^2 \geq 2/\dot{\sigma}_e$, and from \cref{eq:var-exact-sym}, we arrive at
\begin{align}
\label{eq:derivation-1}
\langle\langle\mathcal{J}\rangle\rangle \geq \sum_{e}\frac{2}{\dot{\sigma}_e} \big(d_{B_e}\mathcal{J}\big)^2=\sum_{e} \frac{2}{b_e^2 \dot{\sigma}_e}\big(b_e d_{B_e}\mathcal{J}\big)^2\,.
\end{align}
Sedrakyan's inequality, $\sum_i a_i^2/b_i \geq (\sum_i a_i)^2/(\sum_i b_i)$ that holds for real $a_i$ and positive $b_i$, implies 
\begin{align}
\label{eq:derivation-2}
\sum_{e} \frac{\big(b_e d_{B_e}\mathcal{J}\big)^2}{b_e^2 \dot{\sigma}_e}\geq \frac{ \big( \sum_{e}b_e d_{B_e}\mathcal{J}\big)^2}{\sum_e b_e^2 \dot{\sigma}_e} = \frac{( d_\varepsilon \mathcal{J} )^2}{\sum_e b_e^2 \dot{\sigma}_e} \,,
\end{align}
where we used \cref{eq:responses-eps-eta} for the last equality. 
Noticing that 
\begin{align}
\label{eq:derivation-3}
    \sum_e b_e^2 \dot{\sigma}_e = b_\text{max}^2\dot{\sigma}_\beta \leq b_\text{max}^2\dot{\sigma}_\omega \leq b_\text{max}^2\dot{\sigma}\,,
\end{align}
where $b_\text{max}=\text{max}_e|b_e|$; $\dot{\sigma}_\beta \equiv \sum_e \beta_e^2 \dot{\sigma}_e$ is the weighted EPR with $\beta_e = b_e/b_\text{max}$; $\dot{\sigma}_\omega$ is the partial EPR~\cite{shiraishi2015fluctuation} for the set $\omega$ of perturbed edges (i.e., the set of edges for which $b_e \neq 0$); and where we used $\beta_e^2\leq 1$ for the first inequality. 
\blue{Intuitively, $ b_\text{max}$ characterizes the most abrupt change in barriers resulting from a perturbation in the physical parameter.}
Inserting \cref{eq:responses-eps-eta,eq:derivation-2,eq:derivation-3} in \cref{eq:derivation-1} we find the first hierarchy
\begin{align}
    \label{eq:response-tur-def}
    \frac{( d_\varepsilon\mathcal{J})^2}{b_\text{max}^2\langle\langle\mathcal{J}\rangle\rangle}=
    \frac{(\sum_{e}\beta_e d_{B_e}\mathcal{J})^2}{\langle\langle\mathcal{J}\rangle\rangle}
    \leq
     \frac{\dot{\sigma}_\beta}{2}\leq
     \frac{\dot{\sigma}_\omega}{2}\leq
     \frac{\dot{\sigma}}{2}\,,
\end{align}
which proves and generalizes the R-TUR conjectured in \cite{ptaszynski2024dissipation}.
Using the fact that for single edge perturbation \blue{$\varepsilon = B_e$ and thus $b_\text{max}=1$}, we find
\begin{align}
\label{eq:response-tur-edge}
    \frac{(d_{B_e} \mathcal{J})^2}{\langle\langle\mathcal{J}\rangle\rangle} \leq \frac{\dot{\sigma}_e}{2}\,,
\end{align}
which bounds the response by the single edge EPR.
\blue{
Since \cref{eq:response-tur-def} hold for arbitrary $\beta_e$, using $\beta_e = \text{sign}\,d_{B_e}\mathcal{J}$ for the last inequality in \cref{eq:response-tur-def}, we also find the bound
\begin{align}
\label{eq:tightest-TUR}
     \frac{(\sum_{e} | d_{B_e}\mathcal{J}|)^2}{\langle\langle\mathcal{J}\rangle\rangle}\leq\frac{\dot{\sigma}}{2}\,,
\end{align}
which is tighter than the bound derived from the edge summation of \cref{eq:response-tur-edge}.
}

To derive the second hierarchy, we start from \cref{eq:var-exact-antisym}
\begin{align}
    \label{eq:derivation-antisym-1}
    \frac{\langle\langle\mathcal{J}\rangle\rangle }{4}= \sum_{e}\frac{\big(z_e d_{S_e}\mathcal{J}\big)^2}{z_e^2\tau_e}\geq \frac{ \big( \sum_{e}z_e d_{S_e}\mathcal{J}\big)^2}{\sum_e z_e^2 \tau_e} = \frac{( d_\eta \mathcal{J} )^2}{\sum_e z_e^2 \tau_e} \,,
\end{align}
where we used Sedrakyan's inequality and \cref{eq:responses-eps-eta} for the last equality. As for \cref{eq:derivation-3}, we have
\begin{align}
\label{eq:derivation-antisym-2}
    \sum_e z_e^2 \tau_e = z_\text{max}^2\mathcal{T}_\zeta \leq z_\text{max}^2\mathcal{T}_\varphi \leq z_\text{max}^2\mathcal{T}\,,
\end{align}
which, with \cref{eq:derivation-antisym-1}, results in the \blue{R-KUR} hierarchy
\begin{align}
    \label{eq:traffic-tur-def}
    \frac{( d_\eta\mathcal{J})^2}{z_\text{max}^2\langle\langle\mathcal{J}\rangle\rangle}=
    \frac{(\sum_{e}\zeta_e d_{S_e}\mathcal{J})^2}{\langle\langle\mathcal{J}\rangle\rangle}
    \leq
     \frac{\mathcal{T}_\zeta}{4}\leq
     \frac{\mathcal{T}_\varphi}{4}\leq
     \frac{\mathcal{T}}{4}\,,
\end{align}
where $z_\text{max}=\text{max}_e|z_e|$; $\mathcal{T}_\zeta = \sum_e \zeta_e^2\tau_e$ is the weighted traffic with $\zeta_e =z_e/z_\text{max}$; and $\mathcal{T}_\varphi = \sum_{e\in\varphi}\tau_e$ for the set $\varphi$ defining the perturbed edges ($z_e\neq 0$).
\blue{
We note that the hierarchy of bounds in \cref{eq:traffic-tur-def} can be alternatively obtained using information geometry approach~\cite{zheng2024information}.
}
\blue{
Since $\eta=S_e$ implies $z_\text{max}=1$, using \cref{eq:traffic-tur-def}, we obtain the single edge R-KUR 
\begin{align}
    \label{eq:traffic-tur-edge}
    \frac{( d_{S_e}\mathcal{J})^2}{\langle\langle\mathcal{J}\rangle\rangle}\leq
     \frac{\tau_e}{4}\,.
\end{align}
Repeating the procedure leading to \cref{eq:tightest-TUR}, the last inequality in \cref{eq:traffic-tur-def} can be rewritten as
\begin{align}
\label{eq:tightest-KUR}
    \frac{(\sum_{e} | d_{S_e}\mathcal{J}|)^2}{\langle\langle\mathcal{J}\rangle\rangle}\leq\frac{\mathcal{T}}{4}\,,
\end{align}
which is tighter than the one derived from the edge summation of \cref{eq:traffic-tur-edge}.
}

\blue{
Introducing the squared precision $\phi =\mathcal{J}^2/\langle\langle\mathcal{J}\rangle\rangle$ and the sensitivity of the current observable $d_{p}\ln \mathcal{J}\equiv d_{p}\mathcal{J}/\mathcal{J}$, the single edge R-TURs, \cref{eq:traffic-tur-edge,eq:response-tur-edge}, take the compact form
\begin{align}
    \label{eq:edge-tur-compact}
    \sigma_e \geq 2(d_{B_e}\ln\mathcal{J})^2\phi\,,\quad  \tau_e \geq 4(d_{S_e}\ln\mathcal{J})^2\phi\,.
\end{align}
These results show that dissipation and traffic impose a trade-off between precision and sensitivity. 
}

\textit{Pseudo-EPRs forms of R-TUR and TUR---}
Multiplying both sides of \cref{eq:var-exact-sym} by $\dot{\Pi}$ and using Cauchy–Schwarz inequality $(\sum_i a_i^2)(\sum_i b_i^2)\geq (\sum_i |a_ib_i|)^2$, we arrive at
\begin{align}
\label{eq:var-bound-CS}
    \dot{\Pi}\langle\langle\mathcal{J}\rangle\rangle&=
    \Big(\sum_e\frac{j_e^2}{\tau_e}\Big)\sum_{e} (d_{B_e}\mathcal{J})^2 \frac{\tau_e}{j_e^2}\geq \Big(\sum_e|d_{B_e}\mathcal{J}|\Big)^2\,.
\end{align}
\blue{We note that since $\dot{\Pi}\leq\dot{\sigma}/2$, the bound in \cref{eq:var-bound-CS} is tighter than \cref{eq:tightest-TUR}.}
Since the response $d_\varepsilon\mathcal{J}$ in \cref{eq:responses-eps-eta} is bounded as
\begin{align}
    \label{eq:response-epsilon}    (d_\varepsilon\mathcal{J})^2=\Big(\sum_e b_e d_{B_e}\mathcal{J}\Big)^2\leq b_\text{max}^2\Big(\sum_e  |d_{B_e}\mathcal{J}|\Big)^2\,,
\end{align}
using \cref{eq:response-epsilon,eq:var-bound-CS} together with $\dot{\Pi}\leq\dot{\sigma}/2$, we find 
\begin{align}
    \label{eq:response-tur-pseudo-EPR}
    \frac{( d_\varepsilon\mathcal{J})^2}{\langle\langle\mathcal{J}\rangle\rangle}
    \leq b_\text{max}^2 \dot{\Pi} \leq \frac{b_\text{max}^2\dot{\sigma}}{2}\,.
\end{align}
By bounding the precision of current response to symmetric perturbations using the pseudo-EPR instead of the EPR, this result constitutes a tighter version of the R-TUR.

To recover the standard TUR, we relax \cref{eq:var-bound-CS} by omitting the absolute value inside summations 
\begin{align}
\label{eq:standard-tur-pseudo-EPR-0} \dot{\Pi}\langle\langle\mathcal{J}\rangle\rangle&\geq\Big(\sum_e d_{B_e}\mathcal{J}\Big)^2=\mathcal{J}^2\;,
\end{align}
where in the last step we used $\sum_e d_{B_e} j_{e'}=j_{e'}$, which is a special case of the summation response relations 
\footnote{using $\partial_{B_e}j_e = j_e$ with Eqs.~(11) and (13) of \cite{aslyamov2024general} we arrive at $\sum_e d_{B_e} j_{e'}=j_{e'}$}. 
From \cref{eq:standard-tur-pseudo-EPR-0}, we find
\begin{align}
\label{eq:standard-tur-pseudo-EPR}
    \frac{\mathcal{J}^2}{\langle\langle\mathcal{J}\rangle\rangle}&\leq \dot{\Pi} \leq \frac{\dot{\sigma}}{2}\,.
\end{align}
For the standard TUR, our derivation of \cref{eq:standard-tur-pseudo-EPR} is complementary to the method \cite{shiraishi2021optimal} based on the Cram\'{e}r--Rao inequality and Fisher information; see also \cite{van2022unified,dechant2022minimum}. 
The second inequalities in \cref{eq:response-tur-pseudo-EPR,eq:standard-tur-pseudo-EPR} are saturated if and only if $j_e^2/\tau_e = \dot{\sigma}_e/2$ for $\forall e$, namely close to equilibrium. It explains the simulations in Ref.~\cite{ptaszynski2024dissipation} observing tight bounds only close to equilibrium, even for optimized TUR and R-TUR. Furthermore, since $j_e\leq \tau_e$, we find that $\dot{\Pi}\leq \mathcal{T}$, which using \cref{eq:response-tur-pseudo-EPR,eq:standard-tur-pseudo-EPR} implies that
\begin{align}
    \frac{( d_\varepsilon\mathcal{J})^2}{\langle\langle\mathcal{J}\rangle\rangle}
    \leq b_\text{max}^2 \mathcal{T}\,,\quad\frac{\mathcal{J}^2}{\langle\langle\mathcal{J}\rangle\rangle}&\leq\mathcal{T}\,.
\end{align}
The right inequality is the kinetic-TUR derived in \cite{di2018kinetic} using the method of \cite{dechant2020fluctuation}. The left inequality constitutes a kinetic version of the R-TUR. 

\begin{figure}
    \centering
    \includegraphics[width=\linewidth]{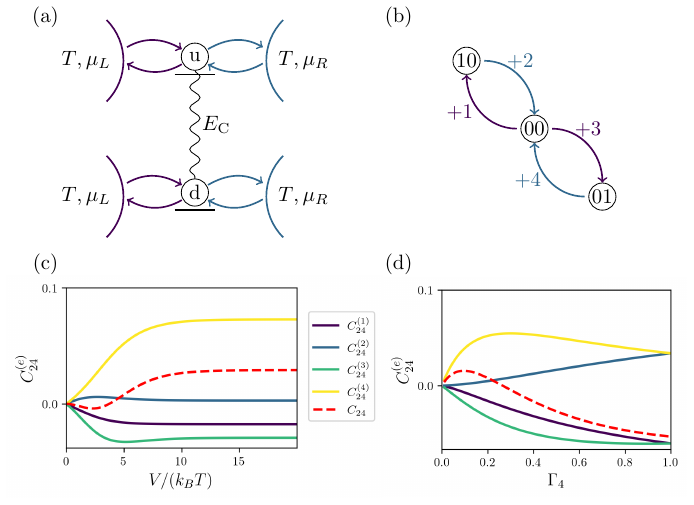}
    \caption{
    (a) The scheme of two QDs, $\text{u}$ and $\text{d}$, (up and down), where every QD is connected to the left and right electrodes ($L$ and $R$) with temperature $T$ and chemical potentials $\mu_L=-\mu_R=V/2$. Purple and blue arrows depict the electron transfer between QDs and the left and right electrodes, respectively. Wavy curve depicts the Coulomb repulsion with energy $E_C$ between filled QDs. 
    (b) The corresponding Markov network in the limit $E_C\gg \{V,k_BT\}$ with three states: $00$, QDs are empty; $10$ or $01$, only the up or down QD is filled. The arrows depict the directed jumps between the states. Purple and blue arrows are assigned to left and right electrodes, respectively. 
    (c) Dashed red line shows the covariance $C_{24} \equiv \langle\langle j_2 ,j_4\rangle\rangle$ as function of the nondimensional potential $V/(k_BT)$ calculated via \cref{eq:cov-example} for $\Gamma_1=\Gamma_2=\Gamma_3=1$ and $\Gamma_4=0.2$. 
    Colored solid curves are four terms from the left side of \cref{eq:cov-example}, where purple, blue, green, yellow correspond to $e\in\{1,2,3,4\}$, respectively. (d) The same as in (c), but functions of $\Gamma_4$ with $V=5 k_B T$. 
    }
    \label{fig:fig-1}
\end{figure}

\textit{Application of FRRs---}\blue{We consider two examples illustrating our approach: the Coulomb-blockaded system based on the quantum dots in the main text and chemical reactions obeying mass action law in \cref{sec:CRN}. 
Both examples show how perturbations in FRRs can be controlled by the physical parameters. 
The quantum dots example focuses on the analysis of the exact FRR. 
In the chemical example, we discuss the inference of EPR and traffic. There, analogously to standard TUR and KUR, we use R-TUR and R-KUR to infer the minimum value of the EPR or of traffic using current fluctuations and responses. 
We illustrate that the EPR and traffic can be inferred using only partial rather than full knowledge of the model.
Furthermore, we derive the bounds for the reaction activity and the ratio of the concentrations of chemical species that have recently attracted attention~\cite{despons2024structural,liang2024thermodynamicBounds,liang2024thermodynamicSpace}.}

In the context of the quantum dot example, we show that FRRs explain the observation, made in~\cite{cottet2004positive,cottet2004positivePRL}, of positive cross-correlations in Coulomb-blockaded systems, which are normally negative.
To do so we consider the dynamical channel blockade model~\cite{bulka2000current,belzig2005full}.
This model consists of two single level Quantum Dots (QD) with energy levels $\epsilon_u=\epsilon_d=0$. The QDs exchange electrons with two electrodes (reservoirs) which have the same temperature $T$ and chemical potentials $\mu_L=-\mu_R=V/2$ creating a bias $V$ for the left to the right; see \cref{fig:fig-1}a.
Electron transitions between the two QDs are impossible, but the two occupied QDs interact with each other through a Coulomb repulsion energy $E_C$. 
We consider the limit of $E_C\gg \{V,k_BT\}$ where the probability of finding both QDs filled is negligible. Thus, the Markov network for this model consists of three states (both QDs are empty, $u$ is filled but not $d$, and vice versa); see the graph in \cref{fig:fig-1}b. The transition rates assigned to the edges $e\in\{1,2,3,4\}$ are $W_{\pm e}=\Gamma_e(1+\exp[\mp V/(2k_BT)])^{-1}$, where $\Gamma_e$ are the tunneling rates. Therefore, the symmetric parameters $B_e=\ln\sqrt{W_{+e}W_{-e}}$ depend on both $\Gamma_e$ and $V$, while the antisymmetric parameters $S_e=\ln(W_{+e}/W_{-e})=V/(2k_BT)$ depend only on $V$. Thus, to perturb only the symmetric parameters, one needs to control the tunneling rates $\Gamma_e$. 
To study the covariance $C_{24} \equiv \langle\langle j_2,j_4\rangle\rangle$, describing correlations between the different currents on the right electrode, we use the FRR \eqref{eq:covar-exact-sym} for symmetric perturbations (parameters $B_e$)
\begin{align}
\label{eq:cov-example}
    C_{24} =\sum_{e=1}^4\frac{\tau_e}{j_e^2}d_{B_e}j_2d_{B_e}j_4=\sum_{e=1}^4C_{24}^{(e)}\,,
\end{align}
where $C_{24}^{(e)}$ are the contributions to the covariance from the edge responses. 
\blue{In Section SIV of \cite{SM} (\cref{sec:matP-QDs}) we derive the following inequalities for the terms in \cref{eq:cov-example},
\begin{align}
    C_{24}^{(1)} \leq 0\,,\,C_{24}^{(2)} \geq 0\,,\,C_{24}^{(3)} \leq 0\,,\,C_{24}^{(4)} \geq 0\,,
\end{align}
which indicate that the positive correlations originate from the
current responses to the perturbation of coupling to the right electrode (the barriers $\{B_2, B_4\}$). 
}
The summation of the terms of different signs in \cref{eq:cov-example,eq:cov-example} makes both positive and negative correlations $C_{24}$ possible.
This can be seen in \cref{fig:fig-1} cd, where we plot the terms of \cref{eq:cov-example} as functions of the bias $V$ in \cref{fig:fig-1}c, and the tunneling rate $\Gamma_4$ in \cref{fig:fig-1}d. 
We see a positive covariance between the currents for sufficiently small rate $\Gamma_4$ and large potential $V$. 

\textit{Conclusions---}FRRs in \cref{eq:covar-exact-sym,eq:var-exact-sym,eq:covar-exact-antisym,eq:var-exact-antisym} reveal a fundamental and general relation between static response and fluctuations giving rise to hierarchies of R-TURs and R-KURs.  
These findings are relevant to any domain utilizing Markov jump processes, as well as the growing body of research on static responses in nonequilibrium systems 
\cite{mitrophanov2005sensitivity,lucarini2016response,santos2020response,falasco2019negative, mallory2020kinetic,owen2020universal,owen2023size,gabriela2023topologically,chun2023trade,aslyamov2024nonequilibrium,aslyamov2024general,harunari2024mutual,floyd2024learning,zheng2024information,frezzato2024steady,floyd2024limits,gao2022thermodynamic,gao2024thermodynamic,khodabandehlou2024affine,harvey2023universal}.
Static response is crucial for understanding transport properties in nanoelectronic devices, such as quantum dots, where it helps characterize responses to changes in tunneling barriers or voltage biases \cite{gustavsson2006counting, gustavsson2009electron, sanchez2019Autonomous}, and in CMOS devices subject to shifts in potential barriers or voltage biases \cite{freitas2021stochastic, gopal2022Large}. In biochemistry, static response plays a pivotal role in quantifying processes like proofreading and sensing \cite{murugan2014discriminatory, owen2020universal, owen2023size}, and in analyzing how metabolic fluxes shift due to changes in enzyme concentrations, as seen in metabolic control analysis \cite{cornish2013fundamentals}, or to changes in thermodynamic forces \cite{falasco2019negative}. Furthermore, static response is instrumental in optimizing loss functions in nonequilibrium physical computing \cite{floyd2024learning, floyd2024limits}.
\blue{
Future studies may concern the application of FRRs in the context of EPR and traffic inference~\cite{seifert2019stochastic,maes2020frenesy}, non-Markovian processes \cite{van2020generalized,ertel2022operationally}, dynamic responses~\cite{mitrophanov2003stability, zheng2025spatial}, and continuous-space Langevin dynamics~\cite{gao2022thermodynamic,gao2024thermodynamic}
}.

\blue{\textit{Author's note---}Before our submission, the authors of \cite{zheng2024information} used information theory to derive the R-KUR bound for arbitrary trajectory observables (including current observables) at the transient regime. 
However, this result does not provide exact FRRs and does not cover the R-TUR.
After submission of our letter, several preprints related to the present work have appeared.
On the one hand, at the level of static response, we extended FRRs to state-based observables in~\cite{ptaszynski2024frr}, and the authors of~\cite{bao2024nonequilibrium} generalized this FRR to ``strong perturbations''.
On the other hand, R-KUR has been generalized to discrete-time Markov processes~\cite{liu2024dynamical} or open quantum systems~\cite{kwon2024fluctuation, van2024fundamental, liu2025response}.
Additionally, Ref.~\cite{kwon2024fluctuation} provided an alternative proof of R-TUR [our Eq.~(15)], and generalized it to state observables, as well as to transient regimes. 
We emphasize that the fluctuation-response relations (FRRs) for covariance, which constitute the primary contribution of our work, are identities that cannot be obtained using the approaches of Refs.~\cite{zheng2024information,liu2024dynamical,kwon2024fluctuation, van2024fundamental, liu2025response}, which are based on the Cram\'{e}r–Rao inequality that bounds covariances. 
}

\begin{acknowledgments}

T.A., K.P.  and M.E. acknowledge the financial support from, respectively, 
project ThermoElectroChem (C23/MS/18060819) from Fonds National de la Recherche-FNR, Luxembourg, 
project No.\ 2023/51/D/ST3/01203 funded by the National Science Centre, Poland, 
project TheCirco (INTER/FNRS/20/15074473) funded by FRS-FNRS (Belgium) and FNR (Luxembourg).
\end{acknowledgments}

\newpage
\appendix

\section{Derivation of covariance matrix and FRRs}
\label{sec:derivation_FRR}

Here we find an exact expression for the covariance matrix $\mathbb{C}$ and then use it to derive the FRRs shown in \cref{eq:covar-exact-sym,eq:covar-exact-antisym,eq:var-exact-sym,eq:var-exact-antisym}.
To find $\mathbb{C}$, we use the scaled cumulant generating function method
\cite{lebowitz1999gallavotti,andrieux2007fluctuation,wachtel2015fluctuating} and define the ``tilted'' matrix $\mathbb{W}^\phi(\boldsymbol{q})$ with nondiagonal elements 
$W_{ij}^\phi(\boldsymbol{q}) = \sum_{e} [W_{+e}\exp(q_e)\delta_{s(+e)j}\delta_{t(+e)i} + W_{-e}\exp(-q_e)\delta_{s(+e)i}\delta_{t(+e)j}]$ 
and diagonal elements the same as $\mathbb{W}$. It implies that $\mathbb{W}^\phi(\boldsymbol{0})=\mathbb{W}$ is the rate matrix.
We calculate the elements of $\mathbb{C}$ as
\begin{align}
\label{eq:appendix-cov-mat}
    C_{ee'} = \frac{\partial}{\partial q_e}\frac{\partial}{\partial q_{e'}} \lambda(\boldsymbol{q})\Big|_{\boldsymbol{q}=\boldsymbol{0}}\,,
\end{align}
where $\lambda(\boldsymbol{q})$ is eigenvalue of the matrix $\mathbb{W}^\phi(\boldsymbol{q})$ with the largest real part.
In \cite{SM}, section SI (\cref{sec:proof-FRR}), we calculate \cref{eq:appendix-cov-mat} to derive an exact expression
\begin{align}
\label{eq:matC}
    \mathbb{C} = \mathbb{P}\mathbb{T}\mathbb{P}^\intercal\,,
\end{align}
with
\begin{align}
    \label{eq:matP}
    \mathbb{P} = \mathbb{1} - \hat{\mathbb{\Gamma}}(\hat{\mathbb{M}}\hat{\mathbb{\Gamma}})^{-1}\hat{\mathbb{M}}\,,
\end{align}
where the matrices $\hat{\mathbb{M}}$ and $\hat{\mathbb{\Gamma}}$ are defined for the oriented graph $\mathcal{G}$ describing the transitions (edges) $\mathcal{E}$ between the states (nodes) $\mathcal{S}$. Then, 
$\hat{\mathbb{M}} = [\delta_{nt(+e)} - \delta_{ns(+e)}]_{\{n\neq N, \forall e\}}$ is the (reduced) incidence matrix of the graph $\mathcal{G}$ with the removed $N$th node; and $\hat{\mathbb{\Gamma}}=[\Gamma_{en}-\Gamma_{eN}]_{\{\forall e,n \neq N\}}$ is defined by the scaled incidence matrix $\Gamma_{en}=W_{+e}\delta_{ns(+e)}-W_{-e}\delta_{nt(+e)}$. The matrix $\hat{\mathbb{M}}\hat{\mathbb{\Gamma}}$ is invertible; see proof in \cite{aslyamov2024general}; the matrix $\mathbb{P}$ is the projection matrix $\mathbb{P}^2=\mathbb{P}$ which was studied in the context of the response theory \cite{aslyamov2024general}.

Finally, we derive FRRs. The key object here is the matrix $\mathbb{P}$ that also defines the static response of $\mathcal{J}$ to perturbations of the arbitrary vector parameter $\boldsymbol{p}=(\dots,p_i,\dots)$ controlling the rates $\mathbb{W}(\boldsymbol{p})$. Indeed, following Ref.~\cite{aslyamov2024general}, we find 
\begin{align}
\label{eq:response-general}
    \nabla^\intercal\mathcal{J}= (\dots, d_{p_i}\mathcal{J},\dots) = \boldsymbol{x}^\intercal \mathbb{P}\mathbb{G}\,,
\end{align}
where $\mathbb{G}=[\partial_{p_i}j_e]_{\forall e,i\in\mathcal{P}}$ is the Jacobian defined in terms of the partial derivatives and which are known from the rates model $\mathbb{W}(\boldsymbol{p})$. If the number of parameters equals the number of edges and they are independent, the Jacobian is invertible ($\mathbb{G}^{-1}$ exists). Therefore, using  \cref{eq:matC,eq:response-general,eq:cov-def}, we find
\blue{
\begin{align}
\label{eq:FRR-general-1}    \langle\langle\mathcal{J},\mathcal{J}'\rangle\rangle &= \boldsymbol{x}^\intercal \mathbb{P}\mathbb{G} \mathbb{G}^{-1} \mathbb{T}(\mathbb{G}^{-1})^\intercal (\boldsymbol{x}'\cdot\mathbb{P}\mathbb{G})^{\intercal}\\ \nonumber
    &=\boldsymbol{x}^\intercal\nabla^\intercal_p\boldsymbol{j} \cdot \mathbb{U} \cdot (\boldsymbol{x}'^\intercal\nabla^\intercal_p\boldsymbol{j})^\intercal \,,
\end{align}
where the vector $\boldsymbol{x}^\intercal\nabla^\intercal_p\boldsymbol{j}=(\dots,\sum_{e'}x_{e'}d_{p_e}j_{e'}\dots,)$ and where  
\begin{align}
\label{eq:matM}
     \mathbb{U}&=\mathbb{G}^{-1} \mathbb{T}(\mathbb{G}^{-1})^\intercal\,.
\end{align}
\Cref{eq:FRR-general-1} is the general result at the core of FFRs. 
In general case, the vector $\boldsymbol{x}$ could be arbitrary implying that
\begin{align}
\label{eq:response-for-arbitraty-x}
    d_{p_e} \mathcal{J} &= \sum_{e'}(d_{p_e} x_{e'}) j_{e'} + \sum_{e'}x_{e'}d_{p_e} j_{e'}\,,\nonumber\\
    (\boldsymbol{x}^\intercal\nabla^\intercal_p\boldsymbol{j})_e &= d_{p_e}\mathcal{J} - \Delta_e\mathcal{J}\,.
\end{align}
where we introduced $\Delta_{e}\mathcal{J}=\sum_{e'}(d_{p_e} x_{e'}) j_{e'}$ the response due to $d_{p_e}\boldsymbol{x}$. 
Inserting \cref{eq:response-for-arbitraty-x} into \cref{eq:FRR-general-1}, we find the general form of FRR as 
\begin{align}
\label{eq:FRR-general-2}
\langle\langle\mathcal{J},\mathcal{J}'\rangle\rangle=\sum_{e,e'}U_{ee'}(d_{p_e}\mathcal{J} - \Delta_{e}\mathcal{J})(d_{p_{e'}}\mathcal{J}' - \Delta_{e'}\mathcal{J}')
\end{align}
\\
In the main text we assume that $\boldsymbol{x}$ does not depend on $\boldsymbol{p}$. Using $d_{p_e} \boldsymbol{x}=0$ in \cref{eq:FRR-general-2} we arrive at
\begin{align}
   \langle\langle\mathcal{J},\mathcal{J}'\rangle\rangle =\nabla^\intercal\mathcal{J}\,\mathbb{U}\, \nabla\mathcal{J}'  \,,
\end{align}
}
If $\boldsymbol{p}=(\dots,B_e,\dots)^\intercal$, we find $\mathbb{G}=\text{diag}(\dots,j_e,\dots)$ and $\mathbb{U} = \mathbb{D}^{-1}$, which gives \cref{eq:covar-exact-sym,eq:var-exact-sym}. 
If $\boldsymbol{p}=(\dots,S_e,\dots)^\intercal$, we find $\mathbb{G}=\mathbb{T}/2$ and $\mathbb{U} = 4\mathbb{T}^{-1}$, which gives \cref{eq:covar-exact-antisym,eq:var-exact-antisym}.

\begin{figure}
    \centering
    \includegraphics[width=0.5\linewidth]{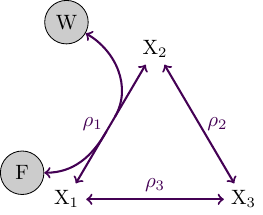}
    \caption{An example of pseudounimolecular reactions $\rho\in\{1,2,3\}$;
    X$_i$ are internal species, $F$ and $W$ are external (chemostatted) species.}
   \label{fig:fig-simple-crn}
\end{figure}

\section{Application to Chemical Reactions}
\label{sec:CRN}

\blue{Here we consider a simple example of chemical reaction networks (CRNs) shown in \cref{fig:fig-simple-crn}, where the network includes three internal chemical species, X$_1$, X$_2$, and X$_3$, interconverted by three reactions. 
One of these reactions is driven by the conversion of an external fuel, F, into waste, W. 
The concentrations of F and W, denoted $c_F$ and $c_W$, respectively, are externally regulated using chemostats. 
In biological terms, F and W could represent ATP and ADP, respectively, and the system in Fig. \ref{fig:fig-simple-crn} could be a simplified model of a molecular motor.}

\blue{Assuming  mass action law (the lack of molecular interaction between species \cite{aslyamov2023nonideal}), the reaction currents read
\begin{align}
\label{eq:CRN-currents}
    \boldsymbol{j}=\big(k_{+1}\pi_1c_F-k_{-1}\pi_2c_W ,\, k_{+2}\pi_2-k_{-2}\pi_3,\, k_{+3}\pi_3-k_{-3}\pi_1\big)^\intercal
\end{align}
where $\pi_i$ is the scaled concentration of the component X$_i$ such that $\sum_i \pi_i = 1$; $k_{\pm\rho}>0$ are reaction constants for $\rho=1, 2, 3$. 
In NESS, the system satisfies
\begin{subequations}
    \begin{align}
\label{eq:CRN-NESS}
    \partial_t \boldsymbol{\pi} &= \mathbb{W}\boldsymbol{\pi} = 0\,,\\
\label{eq:CRN-currents-NESS}
\mathcal{J}&=j_\rho \neq 0\,,
\end{align}
\end{subequations}
with
\begin{align}
\label{eq:CRN-matW}
\mathbb{W}&=
        \begin{pmatrix}
        -k_{+1}c_F-k_{-3} & k_{-1}c_W & k_{+3} \\
        k_{+1}c_F & -k_{-1}c_W- k_{+2} & k_{-2}  \\
        k_{-3} & k_{+2} & -k_{-2}-k_{+3}
        \end{pmatrix}\,,
\end{align}
which plays the role of the rate matrix for the Markov jump processes corresponding to \cref{fig:fig-simple-crn}. }

\blue{The symmetric and asymmetric parts of transition rate $W_{+1}$ can be then expressed as $B_1=\ln \sqrt{k_{+1} k_{-1} c_F c_W}$ and $S_1=\ln[k_{+1} c_F/(k_{-1} c_W)]$. Consequently, using chain rule for derivatives, the response to symmetric and antisymmetric perturbations of the reaction $\rho=1$ can be expressed explicitly in terms of current responses to concentrations (cf.\ Ref.~\cite{owen2020universal}),
\begin{subequations} \label{eq:chain-rule-resp}
\begin{align}
\frac{d \mathcal{J}}{dB_1} &= c_F \frac{d \mathcal{J}}{dc_F} + c_W \frac{d \mathcal{J}}{d c_W }\,, \\
\frac{d \mathcal{J}}{dS_1} &=\frac{1}{2}\left(c_F \frac{d \mathcal{J}}{dc_F}-c_W \frac{d \mathcal{J}}{d c_W }\right) \,.
\end{align}
\end{subequations}
Inserting the above expressions into \cref{eq:response-tur-edge,eq:traffic-tur-edge}, one can bound the entropy production and traffic of reaction $\rho=1$, $\sigma_1$ and $\tau_1$, measuring the current fluctuations $\langle\langle\mathcal{J}\rangle\rangle$ and empirical responses to concentrations $c_F$ and $c_W$,
\begin{subequations}
\label{eq:inference}
\begin{align}
\label{eq:inference-epr}
    \dot{\sigma}_1 \geq  &\frac{2}{\langle\langle\mathcal{J}\rangle\rangle}\Big(c_F \frac{d \mathcal{J}}{dc_F} + c_W \frac{d \mathcal{J}}{d c_W }\Big)^2 \,,\\
\label{eq:inference-tau}
    \tau_1 \geq &\frac{1}{\langle\langle\mathcal{J}\rangle\rangle}\Big(c_F \frac{d \mathcal{J}}{dc_F} - c_W \frac{d \mathcal{J}}{d c_W }\Big)^2\,.
\end{align}
\end{subequations}
We emphasize that we only used the network topology and mass action law, and did not need to know the reaction constants $k_{\pm \rho}$.}

\blue{In the case where the current $\mathcal{J}$ is known, \cref{eq:inference-epr} can be used to bound the edge affinity $\mathcal{A}_1 = \dot{\sigma}_1/\mathcal{J}$. Moreover, from Eqs.~\eqref{eq:CRN-currents}, \eqref{eq:CRN-currents-NESS} and $\tau_1=k_{+1}\pi_1c_F+k_{-1}\pi_2c_W$ we further have $\tau_1=2 c_F \pi_1 k_{+1}-\mathcal{J}$ and $\tau_1=2 c_W \pi_2 k_{-1}+\mathcal{J}$. Inserting those expression into \cref{eq:traffic-tur-edge} and using $\pi_i \leq 1$, we obtain bounds for reaction constants $k_{\pm 1}$:
\begin{subequations}
    \begin{align}
    k_{+1} &\geq \frac{1}{ 2c_F} \left[\frac{4}{\langle\langle \mathcal{J}  \rangle\rangle}\Big(\frac{d \mathcal{J}}{d S_1}\Big)^2+\mathcal{J} \right] \,, \\
     k_{-1} &\geq \frac{1}{2 c_W} \left[\frac{4}{\langle\langle \mathcal{J}  \rangle\rangle}\Big(\frac{d \mathcal{J}}{d S_1}\Big)^2-\mathcal{J} \right] \,.
    \end{align}
\end{subequations}
If the ratio $k_{+1}/k_{-1}$ is known, one can bound the ratio of concentrations $\pi_2/\pi_1$ using the relation $\mathcal{A}_1 = \ln(k_{+1}c_F\pi_1/k_{-1}c_W\pi_2)$.
Indeed, for the graph orientation with $\mathcal{J}>0$, we find
\begin{align}
    \frac{\pi_1}{\pi_2}\geq \frac{k_{-1} c_W}{k_{+1}c_F}\exp\Big[\frac{2}{\mathcal{J}\langle\langle\mathcal{J}\rangle\rangle}\Big(c_F \frac{d \mathcal{J}}{dc_F} + c_W \frac{d \mathcal{J}}{d c_W }\Big)^2\Big]\,.
\end{align}}


\input{sections_for_SM/proof-of-A2}

\input{sections_for_SM/FRR-stalling}
\input{sections_for_SM/QD_example}
\input{sections_for_SM/cycles_of_C}

\bibliography{biblio}
\end{document}

%% file: sections_for_SM/proof-of-A2.tex
\section{Derivation of Eq.~(A2)}
\label{sec:proof-FRR}

Here we derive Eq.~(A2) for the covariance matrix $\mathbb{C}$. To do so, we calculate Eq.~(A1), where the eigenvalue $\lambda(\boldsymbol{q})$ and corresponding eigenvector $\boldsymbol{v}(\boldsymbol{q})$ satisfy
\begin{align}
    \label{eq:eigenval-1}
    \mathbb{W}^\phi(\boldsymbol{q})\boldsymbol{v}(\boldsymbol{q}) = \lambda(\boldsymbol{q})\boldsymbol{v}(\boldsymbol{q})\,,
\end{align}
where $\boldsymbol{v}$ is defined up to normalization and $\lambda(\boldsymbol{0})=0$ with $\boldsymbol{v}(\boldsymbol{0})=\boldsymbol{\pi}$, that allow us to put $\sum_{n}v_n(\boldsymbol{q}) = 1$, which implies
\begin{align}
    \label{eq:eivenvec-response-norm}
    \sum_{n}\partial_{q_e} v_n(\boldsymbol{q}) = 0\,,
\end{align}
for all $\boldsymbol{q}$. Calculating the derivative of \cref{eq:eigenval-1} with respect to the parameter $q_e$ we arrive at
\begin{align}
    \label{eq:eigenval-2}
    (\partial_{q_e}\mathbb{W}^\phi)\boldsymbol{v} + \mathbb{W}^\phi\partial_{q_e}\boldsymbol{v} = (\partial_{q_e}\lambda)\boldsymbol{v}+ \lambda\partial_{q_e}\boldsymbol{v}\,,
\end{align}
which at $\boldsymbol{q}=\boldsymbol{0}$ becomes 
\begin{align}
    \label{eq:eigenval-3}
    (\partial_{q_e}\mathbb{W}^\phi(\boldsymbol{0}))\boldsymbol{\pi} + \mathbb{W}\partial_{q_e}\boldsymbol{v}(\boldsymbol{0}) = (\partial_{q_e}\lambda(\boldsymbol{0}))\boldsymbol{\pi}\,,
\end{align}
where we used $\lambda(\boldsymbol{0})=0$, $\boldsymbol{v}(\boldsymbol{0})=\boldsymbol{\pi}$, $\mathbb{W}^\phi(\boldsymbol{0})=\mathbb{W}$ and all derivatives are calculated at $\boldsymbol{q}=\boldsymbol{0}$.
We notice that the derivative,
\begin{align}
\label{eq:dWde}
\partial_{q_e}\mathbb{W}^\phi(\boldsymbol{q})&=
\begin{blockarray}{ccccccccc}
& &  \color{gray} \dots & \color{gray}t(+e) & \color{gray} \dots & \color{gray} s(+e) & \color{gray}\dots & \\
\begin{block}{cc (ccccccc)}
\color{gray} \vdots & &  \phantom{0} & \phantom{0} & \phantom{0} & \phantom{0} & \phantom{0} & \phantom{0} \\
\color{gray} t(+e) & &  \phantom{0} & \phantom{0} & \phantom{0} & W_{+e}\text{e}^{q_e} & \phantom{0} & \phantom{0} \\
\color{gray} \vdots & &  \phantom{0} & \phantom{0} & \phantom{0} & \phantom{0} & \phantom{0} & \phantom{0} \\
\color{gray} s(+e) & & \phantom{0} & -W_{-e}\text{e}^{-q_e} & \phantom{0} & \phantom{0} & \phantom{0} & \phantom{0} \\
\color{gray} \vdots & &  \phantom{0} & \phantom{0} & \phantom{0} & \phantom{0} & \phantom{0} & \phantom{0}\\
\end{block}
\end{blockarray}\,\,,
\end{align}
has only two nonzero elements, that results in
\begin{subequations}
\label{eq:dWde-2}
    \begin{align}
\label{eq:dWde-j}
    \boldsymbol{1}^\intercal (\partial_{q_e}\mathbb{W}(\boldsymbol{0})) \boldsymbol{\pi} &= j_e\,, \\
\label{eq:dWde-tau}
    \boldsymbol{1}^\intercal (\partial^2_{q_e}\mathbb{W}(\boldsymbol{0})) \boldsymbol{\pi} &= \tau_e\,, 
\end{align}
\end{subequations}
where $\boldsymbol{1}=(\dots,1,\dots)^\intercal$ is the vector of ones. Multiplying both sides of \cref{eq:eigenval-3} by $\boldsymbol{1}^\intercal$ and using \cref{eq:dWde-j} and $\boldsymbol{1}^\intercal\mathbb{W}=\boldsymbol{0}$, we find
\begin{align}
\label{eq:dqlambda}
    \partial_{q_e}\lambda(\boldsymbol{0}) = j_e\,,
\end{align}
which confirms the well known result for the first cumulant (average) of the edge currents \cite{wachtel2015fluctuating}.  

We proceed with the second cumulants and calculate the derivative of \cref{eq:eigenval-2} with respect to $q_{e'}$
\begin{align}
    \label{eq:eigenval-4}
    &(\delta_{ee'}\partial_{q_e}^2\mathbb{W}^\phi)\boldsymbol{v}+(\partial_{q_e}\mathbb{W}^\phi)(\partial_{q_{e'}}\boldsymbol{v}) + (\partial_{q_{e'}}\mathbb{W}^\phi)\partial_{q_e}\boldsymbol{v}+\mathbb{W}^\phi(\partial_{q_{e'}}\partial_{q_e}\boldsymbol{v})  \nonumber\\
    &=(\partial_{q_{e'}}\partial_{q_e}\lambda)\boldsymbol{v}+(\partial_{q_e}\lambda)(\partial_{q_{e'}}\boldsymbol{v})+ (\partial_{q_{e'}}\lambda)\partial_{q_e}\boldsymbol{v}+\lambda(\partial_{q_{e'}}\partial_{q_e}\boldsymbol{v})\,,
\end{align}
where we used $\partial_{q_{e'}}\partial_{q_e}\mathbb{W}^\phi(\boldsymbol{q}) = 0$ for $e\neq e'$ [see \cref{eq:dWde}]. At $\boldsymbol{q}=0$, we have
\begin{align}
    \label{eq:eigenval-5}
    &(\delta_{ee'}\partial^2_{q_e}\mathbb{W}^\phi)\boldsymbol{\pi}+(\partial_{q_e}\mathbb{W}^\phi)\partial_{q_{e'}}\boldsymbol{v} + (\partial_{q_{e'}}\mathbb{W}^\phi)\partial_{q_e}\boldsymbol{v}+\mathbb{W}\partial_{q_{e'}}\partial_{q_e}\boldsymbol{v} = \nonumber\\
    &(\partial_{q_{e'}}\partial_{q_e}\lambda)\boldsymbol{\pi}+j_e(\partial_{q_{e'}}\boldsymbol{v})+ j_{e'}\partial_{q_e}\boldsymbol{v}\,,
\end{align}
where we used $\lambda(\boldsymbol{0})=0$, $\boldsymbol{v}(\boldsymbol{0})=\boldsymbol{\pi}$, $\mathbb{W}^\phi(\boldsymbol{0})=\mathbb{W}$ and \cref{eq:dqlambda};
here and below we omit the notation $(\boldsymbol{0})$ in the derivatives. 
Multiplying both sides of \cref{eq:eigenval-5} by $\boldsymbol{1}^\intercal$ and noticing $\boldsymbol{1}^\intercal\partial_{q_e}\boldsymbol{v}=0$ for $\forall e$ [see \cref{eq:eivenvec-response-norm}] and $\boldsymbol{1}^{\intercal}\mathbb{W}=\boldsymbol{0}$, we obtain
\begin{align}
\partial_{q_{e'}}\partial_{q_e}\lambda = \delta_{ee'}\tau_e+\boldsymbol{1}^\intercal(\partial_{q_e}\mathbb{W}^\phi)\partial_{q_{e'}}\boldsymbol{v} + \boldsymbol{1}^\intercal(\partial_{q_{e'}}\mathbb{W}^\phi)\partial_{q_e}\boldsymbol{v}\,,
\end{align}
where for the first term see \cref{eq:dWde-tau}. 
From \cref{eq:dWde}, we find
\begin{align}
\label{eq:dWde-vector}
\boldsymbol{1}^\intercal\partial_{q_e}\mathbb{W}^\phi(\boldsymbol{q})&=
\begin{blockarray}{ccccccc}
 \color{gray} \dots & \color{gray}t(+e) & \color{gray} \dots & \color{gray} s(+e) & \color{gray}\dots & \\
\begin{block}{(ccccccc)}
    \phantom{0} &  -W_{-e}\text{e}^{-q_e} & \phantom{0} &  W_{+e}\text{e}^{q_e} & \phantom{0} & \phantom{0} \\
\end{block}
\end{blockarray}\,\,,
\end{align}
and at $\boldsymbol{q}=\boldsymbol{0}$, we have
\begin{align}
\label{eq:eigenval-6}
(\boldsymbol{1}^\intercal\partial_{q_e}\mathbb{W}^\phi)_n&=-W_{-e}\delta_{nt(+e)}+W_{+e}\delta_{ns(+e)} = \Gamma_{e n}\nonumber\\
\boldsymbol{1}^\intercal(\partial_{q_e}\mathbb{W}^\phi)\partial_{q_{e'}}\boldsymbol{v} &= \sum_{n=1}^N\Gamma_{en}\partial_{q_{e'}}v_n = \sum_{n\neq N}\Gamma_{en}\partial_{q_{e'}}v_n -\Gamma_{eN}\sum_{n\neq N}\partial_{q_{e'}}v_n \nonumber\\
    &=\sum_{n\neq N}\hat{\Gamma}_{en}\partial_{q_{e'}}v_n\,, 
\end{align}
with
\begin{subequations}
\label{eq:matG}
    \begin{align}
    \Gamma_{en} &= W_{+e}\delta_{n s(+e)} - W_{-e}\delta_{n t(+e)}\,,\\
    \hat{\Gamma}_{en} & = \Gamma_{en} - \Gamma_{e N}\,,
\end{align}
\end{subequations}
where we used $\partial_{q_e}v_N = -\sum_{n\neq N}\partial_{q_e}v_n$ [\cref{eq:eivenvec-response-norm}]. The matrix $\mathbb{\Gamma} = [\Gamma_{en}]_{\{\forall e, \forall n\}}$ is the weighted incidence matrix of the oriented graph $\mathcal{G}$ with the nodes $\mathcal{S}$ and edges $\mathcal{E}$ corresponding the states and transitions of the Markov processes, respectively; 
the matrix $\hat{\mathbb{\Gamma}} = [\Gamma_{en}-\Gamma_{e N}]_{\{\forall e, n\neq N\}}$ is the reduced (column rank) matrix. Thus, using Eq.~(A1),~\cref{eq:eigenval-5,eq:eigenval-6,eq:matG}, we find
\begin{align}
\label{eq:matC-1}
    \mathbb{C} = \mathbb{T} + \hat{\mathbb{\Gamma}}\mathbb{R} + (\hat{\mathbb{\Gamma}}\mathbb{R})^\intercal\,,
\end{align}
where $\mathbb{R}=[\partial_{q_e}v_n]_{\{n\neq N, \forall e\}}$ is the reduced response matrix of the eigenvector $\boldsymbol{v}$ to the parameter vector $\boldsymbol{q}$.

We proceed with calculations of the matrix $\mathbb{R}$.
Using \cref{eq:eivenvec-response-norm}, we have $\sum_{m}W_{nm}\partial_{q_e}v_m=\sum_{m\neq N}(W_{nm}-W_{nN})\partial_{q_e}v_m=\sum_{m\neq N}K_{nm}R_{me}$. Inserting this relation into \cref{eq:eigenval-3} and solving it for $\mathbb{R}$, we arrive at
\begin{align}
\label{eq:matR}
    \sum_{m\neq N}K_{nm}R_{me} &= \pi_n j_e - \sum_{m=1}^N\partial_{q_e}W_{nm}^\phi\pi_m\,,\quad\text{for}\,n\neq N\,,\nonumber\\
    \mathbb{R} &= \mathbb{K}^{-1}(\hat{\boldsymbol{\pi}} \otimes \boldsymbol{j} - \mathbb{Z})\,,
\end{align}
where $\hat{\boldsymbol{\pi}}=(\pi_1,\dots,\pi_{N-1})^{\intercal}$, and $\otimes$ denoted the outer product $\hat{\pi} \otimes j = [\pi_n j_e]_{\{n\neq N,\forall e\}}$, and where we also introduced
\begin{align}
\label{eq:matZ-1}
    \mathbb{Z} &= \Big[\sum_{m}\partial_{q_e}W^\phi_{nm}\pi_m\Big]_{\{n\neq N, \forall e\}}\nonumber\\
    &=\Big[W_{+e}\pi_{s(+e)}\delta_{nt(+e)}-W_{-e}\pi_{t(+e)}\delta_{ns(+e)}\Big]_{\{n\neq N, \forall e\}}\nonumber\\
    &=\frac{1}{2}\Big[M_{ne}\tau_e+Q_{ne}j_e\Big]_{\{n\neq N, \forall e\}}=\frac{1}{2}(\hat{\mathbb{M}}\mathbb{T}+\hat{\mathbb{Q}}\mathbb{J})\,,
\end{align}
and
\begin{align}
\label{eq:matK}
    \mathbb{K} = [W_{nm}-W_{nN}]_{\{n\neq N, m \neq N\}} = \hat{\mathbb{M}}\hat{\mathbb{\Gamma}}\,, 
\end{align}
where $\mathbb{J}=\text{diag}(\dots,j_e,\dots)$, $\hat{\mathbb{M}} = [\delta_{nt(+e)} - \delta_{ns(+e)}]_{\{n\neq N, \forall e\}}$ is the reduced incidence matrix of the oriented graph $\mathcal{G}$, and similarly $\hat{\mathbb{Q}} = [\delta_{ns(+e)}+\delta_{nt(+e)}]_{\{n\neq N, \forall e\}}$ is defined for the non-oriented graph with the same topology as $\mathcal{G}$. 
The matrix $\mathbb{K}$ is invertible; see the proof $\det\mathbb{K}\neq0$ in~\cite{aslyamov2024general}. 

In terms of the incidence matrix $\mathbb{M}$, the steady state must satisfy $\mathbb{M}\boldsymbol{j}=\boldsymbol{0}$, which is equivalent to $\hat{\mathbb{M}}\boldsymbol{j}=\boldsymbol{0}$. 
This property transfers to the covariance matrix $\hat{\mathbb{M}}\mathbb{C} = 0$; see the analysis of the Schnakenberg cycles of the covariance matrix in Refs.~\cite{wachtel2015fluctuating,polettini2016tightening} and an alternative derivation in \cref{sec:cycles_matC}.
Multiplying \cref{eq:matC-1} by $\hat{\mathbb{M}}$ and using \cref{eq:matR,eq:matK}, we find
\begin{align}
\label{eq:matZ-2}
    \hat{\mathbb{M}}\mathbb{T} + \underbrace{\hat{\mathbb{M}}\hat{\mathbb{\Gamma}}\mathbb{K}^{-1}}_{\mathbb{1}}&(\hat{\boldsymbol{\pi}} \otimes \boldsymbol{j} - \mathbb{Z}) + (\underbrace{\hat{\mathbb{M}}\boldsymbol{j}\otimes \hat{\boldsymbol{\pi}}}_{0}  - \hat{\mathbb{M}}\mathbb{Z}^\intercal)(\hat{\mathbb{\Gamma}}\mathbb{K}^{-1})^{\intercal}=0\,,\nonumber\\
    &\hat{\boldsymbol{\pi}} \otimes \boldsymbol{j} - \mathbb{Z} =  \hat{\mathbb{M}}(\hat{\mathbb{\Gamma}}\mathbb{K}^{-1}\mathbb{Z})^{\intercal} - \hat{\mathbb{M}}\mathbb{T}\,,
\end{align}
which we plug in \cref{eq:matR} and using \cref{eq:matZ-1}, we arrive at
\begin{align}
\label{eq:matR-2}
    \mathbb{R} &= \mathbb{K}^{-1}\hat{\mathbb{M}}\big[(\hat{\mathbb{\Gamma}}\mathbb{K}^{-1}\mathbb{Z})^{\intercal} -\mathbb{T}\big]\nonumber\\
    &=\tfrac{1}{2}\mathbb{K}^{-1}\hat{\mathbb{M}}\big[\mathbb{T}(\hat{\mathbb{\Gamma}}\mathbb{K}^{-1}\hat{\mathbb{M}})^{\intercal}+\mathbb{J}(\hat{\mathbb{\Gamma}}\mathbb{K}^{-1}\hat{\mathbb{Q}})^{\intercal} -2\mathbb{T}\big]\nonumber\\
    &=\tfrac{1}{2}\mathbb{K}^{-1}\hat{\mathbb{M}}\big[\mathbb{T}\mathbb{P}_M^{\intercal}+\mathbb{J}\mathbb{P}_Q^{\intercal} -2\mathbb{T}\big]\,,
\end{align}
where we introduced $\mathbb{P}_M=\hat{\mathbb{\Gamma}}\mathbb{K}^{-1}\hat{\mathbb{M}}$ and $\mathbb{P}_Q=\hat{\mathbb{\Gamma}}\mathbb{K}^{-1}\hat{\mathbb{Q}}$. Using \cref{eq:matR-2}, we calculate the second and third terms in \cref{eq:matC-1} as
\begin{subequations}
    \begin{align}
        \hat{\mathbb{\Gamma}}\mathbb{R} &= \frac{1}{2}(\mathbb{P}_M\mathbb{T}\mathbb{P}_M^\intercal+\mathbb{P}_M\mathbb{J}\mathbb{P}_Q^\intercal) - \mathbb{P}_M\mathbb{T} \,,\\
        (\hat{\mathbb{\Gamma}}\mathbb{R})^\intercal &= \frac{1}{2}(\mathbb{P}_M\mathbb{T}\mathbb{P}_M^\intercal+\mathbb{P}_Q\mathbb{J}\mathbb{P}_M^\intercal) - \mathbb{T}\mathbb{P}_M^\intercal\,,
    \end{align}
\end{subequations}
that allow us to rewrite \cref{eq:matC-1} as
\begin{align}
    \label{eq:matC-2}
    \mathbb{C} = \mathbb{P} \mathbb{T} \mathbb{P}^\intercal + \frac{1}{2}(\mathbb{P}_M\mathbb{J}\mathbb{P}_Q^\intercal + \mathbb{P}_Q\mathbb{J}\mathbb{P}_M^\intercal)\,,
\end{align}
where we introduced
\begin{align}
\label{eq:matP}
    \mathbb{P} = \mathbb{1} - \mathbb{P}_M = \mathbb{1} - \hat{\mathbb{\Gamma}}(\hat{\mathbb{M}}\hat{\mathbb{\Gamma}})^{-1}\hat{\mathbb{M}}\,,
\end{align}
which is the projection matrix $\mathbb{P}$; see its properties in \cite{aslyamov2024general}. 

We proceed with the calculation of the second and third terms in \cref{eq:matC-2}. Using the notation $\hat{\mathbb{M}}=\hat{\mathbb{M}}_- - \hat{\mathbb{M}}_+$, $\hat{\mathbb{Q}}=\hat{\mathbb{M}}_+ + \hat{\mathbb{M}}_-$, and $\mathbb{A}=\hat{\mathbb{\Gamma}}\mathbb{K}^{-1}$, we write
\begin{align}
    \label{eq:mat-zero-1}
     \mathbb{P}_M\mathbb{J}\mathbb{P}_Q^\intercal = \mathbb{A} (\hat{\mathbb{M}}_- - \hat{\mathbb{M}}_+) \mathbb{J} (\hat{\mathbb{M}}_+ + \hat{\mathbb{M}}_-)^\intercal\mathbb{A}^\intercal\,,
\end{align}
and then calculate
\begin{align}
    \label{eq:mat-zero-2}
     \mathbb{P}_M\mathbb{J}\mathbb{P}_Q^\intercal+(\mathbb{P}_M\mathbb{J}\mathbb{P}_Q^\intercal)^\intercal = 2\mathbb{A}(\hat{\mathbb{M}}_-\mathbb{J}\hat{\mathbb{M}}_-^\intercal-\hat{\mathbb{M}}_+\mathbb{J}\hat{\mathbb{M}}_+^\intercal)\mathbb{A}^\intercal\,.
\end{align}
We notice that the element $(n,m)$ of the matrix $\hat{\mathbb{M}}_-\mathbb{J}\hat{\mathbb{M}}_-^\intercal-\hat{\mathbb{M}}_+\mathbb{J}\hat{\mathbb{M}}_+^\intercal$ is
\begin{align}
 \label{eq:mat-zero-3}
     &\sum_{e}(\delta_{nt(+e)}j_e\delta_{m t(+e)}-\delta_{ns(+e)}j_e\delta_{m s(+e)})=\nonumber \\
    &\sum_{e} (\delta_{nt(+e)}-\delta_{ns(+e)})\delta_{nm}j_e = 
    \delta_{nm}\sum_e M_{ne} j_e = 0\,.
\end{align}
Thus, all elements of the matrix \cref{eq:mat-zero-2} are zero, which proves that
\begin{align}
\label{eq:matrix_zero}
    \mathbb{P}_M\mathbb{J}\mathbb{P}_Q^\intercal + \mathbb{P}_Q\mathbb{J}\mathbb{P}_M^\intercal = 0\,,
\end{align}
which allows us to find the covariance matrix from \cref{eq:matC-2} as Eq.~(A2).

%% file: sections_for_SM/FRR-stalling.tex
\section{FRR for detailed balanced edges}
\label{sec:FRR-stalling}
We use FRR to derive the fluctuation-dissipation relation at stalling originally derived in \cite{altaner2016fluctuation}; see also the alternative derivation in Refs.~\cite{polettini2019effective,polettini2017effective} and in section 10.1 of Ref.~\cite{shiraishi2023introduction}. We consider two edges $e$ and $e'$ with zero average currents $j_e=0$ and $j_{e'}=0$. To calculate the covariance between these two currents, we use \cref{eq:matC-1} as
\begin{align}
\label{eq:matC-stalling}
    C_{ee'}=\tau_e\delta_{ee'} + \sum_{m\neq N}\hat{\Gamma}_{em}R_{me'}+\sum_{m\neq N}\hat{\Gamma}_{e'm}R_{me}\,.
\end{align}
We find $R_{me}$ (resp. $R_{me'}$) inserting $j_e=0$ (resp. $j_{e'}=0$) into \cref{eq:matR} as
\begin{align}
\label{eq:matR-stalling}
    R_{me} = -\frac{1}{2}(\mathbb{K}^{-1}\hat{\mathbb{M}}\mathbb{T})_{me}\,,
\end{align}
where we used $Z_{ne}=\tfrac{1}{2}M_{ne}\tau_e=\tfrac{1}{2}(\hat{\mathbb{M}}\mathbb{T})_{ne}$ [see \cref{eq:matZ-1} with $j_{e}=0$]. Inserting \cref{eq:matR-stalling} in \cref{eq:matC-stalling}, we find
\begin{align}
\label{eq:cov-stalling-1}
    C_{ee'}&=\tau_e\delta_{ee'} - \frac{1}{2}(\hat{\mathbb{\Gamma}}\,\mathbb{K}^{-1}\hat{\mathbb{M}}\mathbb{T})_{ee'} - \frac{1}{2}(\hat{\mathbb{\Gamma}}\,\mathbb{K}^{-1}\hat{\mathbb{M}}\mathbb{T})_{e'e}\nonumber\\
    &=\tau_e\delta_{ee'} - \frac{1}{2}(\mathbb{P}_M\mathbb{T})_{ee'} - \frac{1}{2}(\mathbb{P}_M\mathbb{T})_{e'e}\nonumber \\
    & = \frac{1}{2}(\mathbb{P}\mathbb{T})_{ee'}+\frac{1}{2}(\mathbb{P}\mathbb{T})_{e'e}\,,
\end{align}
where $\mathbb{P}_M=\hat{\mathbb{\Gamma}}\,\mathbb{K}^{-1}\hat{\mathbb{M}}$ and where we used \cref{eq:matP} in the last equality. Using the expression for responses, $d_{S_{e'}}j_e=P_{ee'}T_{e'e'}/2$, we rewrite \cref{eq:cov-stalling-1} as
\begin{align}
    \label{eq:cov-stalling-2}
    C_{ee'} = \frac{d j_e}{d S_{e'}} + \frac{d j_{e'}}{d S_{e}}\,,
\end{align}
which is equivalent to Eq.~(8) of \cite{altaner2016fluctuation}. Following \cite{altaner2016fluctuation}, we introduce the matrix elements $X_{\alpha e}$ for a general current at stalling state, $\mathcal{J}_\alpha=\sum_{e\in \mathcal{X}}X_{\alpha e}j_e$, where $X_{\alpha e} = 0$ for $j_e\neq 0$; and control parameters $h_\alpha$ satisfying $S_e = \sum_{\alpha}X_{\alpha e}h_\alpha$. Thus, using \cref{eq:cov-stalling-2} we derive FRR for stalling state as
\begin{align}
     \label{eq:var-stalling}
     \langle\langle\mathcal{J}_\alpha\rangle\rangle=\sum_{e,e'}X_{\alpha e}C_{ee'}X_{\alpha e'} = 2\sum_{e'}\frac{d \mathcal{J}_\alpha}{d S_{e'}}\frac{\partial S_{e'}}{\partial h_\alpha} = 2 d_{h_\alpha} \mathcal{J}_\alpha\,,
\end{align}
which is equivalent to Eq.~(10) from Ref.~\cite{altaner2016fluctuation}.

%% file: sections_for_SM/QD_example.tex
\section{Calculations for Eq.~(28)}

We introduce the scaled responses
\begin{align}
    P_{ee'}&=\frac{1}{j_{e'}}\frac{d j_e}{d B_{e'}}\,,
\end{align}
and rewrite Eq.~(27) in the main text as
\begin{align}
\label{eq:cov-SM-1}
    C_{24} &=\sum_{e=1}^4\frac{\tau_e}{j_e^2}d_{B_e}j_2d_{B_e}j_4 = \sum_{e=1}^4\tau_e P_{2e}P_{4e} \,,
\end{align}
where the elements $P_{ee'}$ are defined by the projection matrix $\mathbb{P}$; see \cref{eq:matP}. 
The diagonal elements $P_{22}$ and $P_{44}$ are bounded as $0\leq P_{ee}\leq 1$, see Ref.~\cite{aslyamov2024general}. The nondiagonal elements, $P_{ee'}$ with $e\neq e'$, in general, can be negative or positive. However, as shown in \cite{aslyamov2024general} it is possible to express some of nondiagonal elements as the linear combination of the diagonal ones. Applying this method to the Markov network from Fig.~1b we find
\label{sec:matP-QDs}
\begin{align}
\label{eq:matP-example}
    \mathbb{P} = 
    \begin{pmatrix}
        P_{11} & P_{22} & - y_1 & y_1\\
        P_{11} & P_{22} & - y_1 & y_1 \\
        -y_2 & y_2 & P_{33} & P_{44} \\
        -y_2 & y_2 & P_{33} & P_{44} 
    \end{pmatrix}\,,
\end{align}
where $P_{11}+P_{22}=1$ and $P_{33}+P_{44}=1$; at this moment, the parameters $y_1$ and $y_2$ are known. To find the elements of \cref{eq:matP-example} in terms of $W_{\pm e}$ we notice the property of $\mathbb{P}$ [see \cref{eq:matP}]:
\begin{align}
    \mathbb{P}\hat{\mathbb{\Gamma}} = \hat{\mathbb{\Gamma}} - \hat{\mathbb{\Gamma}}(\hat{\mathbb{M}}\hat{\mathbb{\Gamma}})^{-1}\hat{\mathbb{M}}\hat{\mathbb{\Gamma}}=0\,,
\end{align}
where the elements of $\hat{\mathbb{\Gamma}}$ are defined by the rates $W_{\pm e}$ and graph topology introduced in Fig.~1.  Solving $\sum_{e'}P_{1e'}\hat{\Gamma}_{e'n}=0$ for $n=1,2$ with $P_{11}+P_{22}=1$ we find
\begin{align}
y_1 & = \frac{(-1 + \exp(V)) \Gamma_1 \Gamma_2}{\Phi}\,,  \\
P_{11} & = \frac{\Gamma_2 \left( \Gamma_3 + \exp(V)(\Gamma_3 + \Gamma_4) + \exp\left(\frac{V}{2}\right)(\Gamma_3 + 2 \Gamma_4) \right)}{\Phi}\,,  \\
P_{22} & = \frac{\Gamma_1 \left( \Gamma_3 + \Gamma_4 + \exp(V)\Gamma_4 + \exp\left(\frac{V}{2}\right)(2 \Gamma_3 + \Gamma_4) \right)}{\Phi}\,,
\end{align}
and form equations $\sum_{e'}P_{3e'}\hat{\Gamma}_{e'n}=0$ for $n=1,2$ with $P_{33}+P_{44}=1$ we find
\begin{align}
y_2 & = \frac{(-1 + \exp(V)) \Gamma_3 \Gamma_4}{\Phi}\,,  \\
P_{33} & = \frac{\left( \Gamma_1 + \exp(V)(\Gamma_1 + \Gamma_2) + \exp\left(\frac{V}{2}\right)(\Gamma_1 + 2 \Gamma_2) \right) \Gamma_4}{\Phi}\,,  \\
P_{44} & = \frac{\left( \Gamma_1 + \Gamma_2 + \exp(V) \Gamma_2 + \exp\left(\frac{V}{2}\right)(2 \Gamma_1 + \Gamma_2) \right) \Gamma_3}{\Phi}\,,
\end{align}
where
\begin{align}
    \Phi =& \Gamma_1 \left( \Gamma_3 + \Gamma_4 + \exp(V) \Gamma_4 + \exp\left(\frac{V}{2}\right)(2 \Gamma_3 + \Gamma_4) \right) + \nonumber\\
    &\Gamma_2 \left( \Gamma_3 + \exp(V)(\Gamma_3 + \Gamma_4) + 
    \exp\left(\frac{V}{2}\right)(\Gamma_3 + 2 \Gamma_4) \right)\,.
\end{align}
Thus for $V>0$  we have $y_1\geq 0$ and $y_2\geq 0$, which allows us to find the signs for all elements of $\mathbb{P}$ in \cref{eq:matP-example}.
Using this information we rewrite \cref{eq:cov-SM-1} as


\begin{align}
\label{eq:cov-example-scaled-responses}
    C_{24} &= \underbrace{\tau_1 P_{21}P_{41}}_{C_{24}^{(1)}\leq 0} + \underbrace{ \tau_2 P_{22}P_{42}}_{C^{(2)}_{24}\geq  0} +\underbrace{\tau_3 P_{23}P_{43}}_{C_{24}^{(3)}\leq 0} + \underbrace{\tau_4 P_{24}P_{44}}_{C^{(4)}_{24}\geq  0}\,,
\end{align}
where the signs correspond Eq.~(28) in the main text. Thus, we reveal that the scaled responses to perturbations of the coupling to the left electrode are of opposite signs, $P_{21}P_{41} \leq 0$ and $P_{23}P_{43} \leq 0$, contributing with the negative sign to the covariance. 
Therefore, the positive correlation $C_{24}>0$ originates from the scaled responses to perturbations of the coupling to the right electrode, as they are always positive, $P_{24}\geq 0$ and $P_{42}\geq 0$.


%% file: sections_for_SM/cycles_of_C.tex
\section{Derivation of $\mathbb{M} \mathbb{C}=0$}
\label{sec:cycles_matC}

Here we prove the relation $\mathbb{M} \mathbb{C}=0$, which is equivalent to $(\mathbb{M} \mathbb{C})_{ne}=0$ for $\forall_{n,e}$. To this goal, we calculate the elements $(\mathbb{M} \mathbb{C})_{ne}$ as 
\begin{align} \label{eq:proof-mc0-1} \nonumber
(\mathbb{M} \mathbb{C})_{ne} &= \sum_{e'} M_{ne'} C_{e'e}=\sum_{e'} [\delta_{nt(+e')}-\delta_{ns(+e')}] C_{e'e} \\ &= \lim_{t \rightarrow \infty} t^{-1} \langle \Delta k_{e}(t) \Delta k_n(t) \rangle \,,
\end{align}
where $k_n(t)=\sum_e [\delta_{nt(+e)}-\delta_{ns(+e)}] k_e(t)$ is the number of jumps to the state $n$ minus the number of jumps out of that state, and $\Delta k_n(t)=k_n(t)-\langle k_n(t) \rangle$; here in the last step we used the definition of the covariance matrix elements. Applying Cauchy-Schwarz inequality to \cref{eq:proof-mc0-1} we get
\begin{align} \label{eq:proof-mc0-2} \nonumber
&(\mathbb{M} \mathbb{C})^2_{ne}= \lim_{t \rightarrow \infty} \left( \langle [\Delta k_{e}(t)/\sqrt{t}] [ \Delta k_n(t)/\sqrt{t}] \rangle \right)^2 \\ \nonumber &\leq \lim_{t \rightarrow \infty} t^{-1} \langle [\Delta k_{e}(t)]^2 \rangle \times \lim_{t \rightarrow \infty} t^{-1} \langle [\Delta k_{n}(t)]^2 \rangle \\ &= C_{ee} \lim_{t \rightarrow \infty} t^{-1} \langle [\Delta k_{n}(t)]^2 \rangle \,.
\end{align}
We now note that the random variable $k_n(t)$ can take only three values that are not extensive in time:
\begin{itemize}
\item $k_n(t)=1$ when the state $n$ is initially unoccupied and finally occupied,
\item $k_n(t)=0$ when the state $n$ is both initially and finally occupied or unoccupied,
\item $k_n(t)=-1$ when the state is initially occupied and finally unoccupied.
\end{itemize}
Consequently, $\langle k_n(t) \rangle \in [-1,1]$, $ \Delta k_n(t) \in (-2,2)$, and thus $\lim_{t \rightarrow \infty} t^{-1} \langle [\Delta k_{n}(t)]^2 \rangle=0$. As a result, from \cref{eq:proof-mc0-2} we get $(\mathbb{M} \mathbb{C})^2_{ne} \leq 0 \rightarrow (\mathbb{M} \mathbb{C})_{ne}=0$, which proves our statement.